\documentclass[prl,twocolumn,showpacs,amsmath,amssymb,nofootinbib]{revtex4}

\newcommand\lsim{\lesssim}

\newcommand\vev[1]{{\langle {#1} \rangle}}
\renewcommand\({\left(}
\renewcommand\){\right)}
\renewcommand\[{\left[}
\renewcommand\]{\right]}

\newcommand\eq[1]{Eq.~(\ref{#1})}
\newcommand\eqs[2]{Eqs.~(\ref{#1}) and (\ref{#2})}

\newcommand\ee{\end{equation}}
\newcommand\be{\begin{equation}}
\newcommand\eea{\end{eqnarray}}
\newcommand\bea{\begin{eqnarray}}

\def\calp{{\cal P}}

\newcommand\bfk{{\mathbf k}}

\newcommand\bfp{{\mathbf p}}
\newcommand\bfq{{\mathbf q}}

\newcommand\bfx{{\mathbf x}}

\newcommand\sub[1]{_{\rm #1}}

\newcommand\half{^{1/2}}
\newcommand\third{^{1/3}}
\newcommand\quarter{^{1/4}}

\newcommand{\zetag}{{\zeta\sub g}}
\newcommand\sigmas{{\sigma^2}}
\newcommand\sigmam{ {\vev {\sigma} _M} }

\newcommand{\fnl}{{f\sub{NL}}}

\newcommand{\tnl}{{\tau\sub{NL}}}

\newcommand\kmax{{k\sub{max}}}
\newcommand\bfkp{{{\bfk}'}}

\newcommand\bfpp{{{\bfp}'}}

\newcommand\tpq{{(2\pi)^3}}
\newcommand\tps{{(2\pi)^6}}

\newcommand\ksmooth{{k\sub{max}}}
\newcommand\lmone{{L^{-1}}}
\newcommand\zetasig{{\zeta_\sigma}}
\newcommand\zetas{{\zeta_\sigma}}

\begin{document}

\title{Detecting a small perturbation through its non-Gaussianity}

\author{Lotfi Boubekeur and David H.~Lyth}

\affiliation{{\it Physics Department, Lancaster University, Lancaster
LA1 4YB, UK}}

\begin{abstract}
A highly non-gaussian cosmological perturbation with a flat spectrum
has unusual 
stochastic properties.  We show that they depend on the size of the box 
within which the perturbation is defined, but that for a typical observer
the parameters defining the 
 perturbation `run' to compensate for any change in the box size.
Focusing on  the
primordial curvature perturbation, we  show that
 an un-correlated gaussian-squared component is bounded
at around the $10\%$ level by the WMAP  bound on the bispectrum, and we
show that a competitive 
 bound may follow from the trispectrum when it too
is bounded by  WMAP.
Similar considerations apply to a highly non-gaussian 
isocurvature perturbation.

\end{abstract}
\maketitle

{\it Introduction.}~ 
The origin of structure in the Universe seems to be 
the primordial curvature  perturbation $\zeta$,  present already a few Hubble
times before cosmological scales enter the horizon and come into causal
contact \cite{treview}.  Within the observational uncertainty, $\zeta$ is
Gaussian with a practically scale-independent spectrum.  Future
observation though may find  non-Gaussianity or scale-dependence and in
this Letter we focus on the former.
We shall discount the possible contribution
of topological defects to observables, but
we  shall  consider briefly the possible contribution
of a primordial  isocurvature perturbation.

We assume that $\zeta$   originates from the vacuum
fluctuations during  slow-roll inflation of one or more light scalar fields.
These  fluctuations are  promoted to   practically gaussian \cite{sl}
classical perturbations around the time of  horizon exit.  
Expanding $\zeta$ in powers of the field perturbations, the linear terms
are Gaussian and quadratic terms are expected to account adequately for
non-Gaussianity \cite{lr2}.

To obtain detectable non-Gaussianity some 
field  other than the inflaton  must contribute to $\zeta$  
\cite{sl,maldacena}. Possible
examples of such a field in the literature 
are the extra field in double inflation \cite{double},
the curvaton \cite{curvaton} (see also \cite{also}),
 and a  field causing inhomogeneous reheating \cite{reheating}
 or preheating \cite{preheating}.

When discussing non-Gaussianity, it
 is usually supposed that the  contribution of the extra field
will actually dominate 
$\zeta$. Then $\zeta$ is of the form 
\be \zeta(\bfx)  = \sigma(\bfx)  -\frac35 \fnl \( \sigma^2(\bfx) -
\vev{\sigma^2(\bfx)} \) \label{corrchis} \,, 
\ee 
where $\sigma(\bfx)$ is the perturbation of the extra  field,
normalized so that the linear term has unit coefficient.\footnote
{We are defining the
curvature perturbation as in \cite{sasaki1}, and the factor $-3/5$
appears because the original authors \cite{spergel} worked  (using
first-order perturbation theory) with $\Phi\equiv -\frac35\zeta$.}

We will suppose instead that the additional contribution is
sub-dominant so that 
\bea
\zeta(\bfx) &=&  \zetag(\bfx) + \zetasig(\bfx) \\
\zetasig(\bfx)  &=&  a\sigma(\bfx) + \sigma^2(\bfx) - \vev{\sigma^2}
\label{withlin} \,,
\eea
where $\zetag$ is the practically Gaussian contribution of the
 inflaton, and we now
normalize $\sigma$ so that the quadratic term has unit coefficient.
For the moment we take the linear term to be negligible so that\footnote
{The literature includes examples where the linear term dominates
 \cite{Langlois:2004nn} and where the quadratic term dominates
\cite{ev,lr2}. The most closely related previous
discussion is \cite{Fan:1992wv}, whose authors consider \eq{corrchis}
but with a sharply peaked spectrum for $\sigma$ instead of the usual flat
spectrum.}
\be
\zeta_\sigma(\bfx) =   \sigma^2(\bfx) - \vev{\sigma^2}
\label{uncorrchis} \,.
\ee

{\it Basic Definitions.}~ To describe
the stochastic properties of cosmological perturbations one
formally invokes an ensemble of universes, of which the observable
Universe is supposed to be a typical member. A sampling of the
ensemble may be regarded as a sampling of different locations for
the region under consideration. 
The stochastic properties are
conveniently described using a Fourier expansion, which 
we make inside a finite box of coordinate size $L$ much bigger than the region
of interest. In terms of physically significant wavenumbers, this means
$k\gg \lmone$.

We will denote a generic cosmological perturbation, evaluated at some 
instant, by $g(\bfx)$, and 
assume  $\vev g=0$. 
The fields responsible for $\zeta$ are smoothed on
some scale $k_{\rm max}^{-1}$ and we consider the era when this scale 
is outside the horizon. Taking the box size to be formally infinite, the
 spectrum $P_g(k)$ is 
 defined by
\be
\vev{g_\bfk\, g_\bfkp}= \tpq \delta^{(3)}(\bfk+\bfkp ) P_g(k)
\,,
\ee
where  $g_\bfk =  \int d^3x\, e^{i\bfk\cdot\bfx} \,g(\bfx) $ are the Fourier
modes of $g(\bfx)$.
The $\delta$-function and the dependence of  the spectrum
only on $k\equiv |\bfk|$ express the fact that stochastic properties are
invariant under rotations and translations. 
The  variance is  
\be
\vev{ g^2(\bfx) } = \frac1\tpq \int d^3k\, P_g(k) \equiv
\int^\ksmooth_\lmone \frac{dk}{k} \,\calp_g(k)
\,,
\ee
and $\calp_g \equiv (k^3/2\pi^2) P_g$ is the typical value of $g^2$. 
On cosmological scales  $\calp_\zeta$ is almost scale-invariant with
$\calp_\zeta\half =5\times 10^{-5}$.

Non-Gaussianity is signaled by additional connected correlators. 
The bispectrum $B_g$ is defined by \be \vev{ g_{\bfk_1} g_{\bfk_2}
g_{\bfk_3} } = \tpq \delta^{(3)}(\bfk_1+\bfk_2+\bfk_3)
B_g(k_1,k_2,k_3) 
\,.
\ee
If the curvature perturbation has  the form
\eqref{corrchis}  its bispectrum is given to leading order by \cite{spergel}
\be
 B_\zeta(k_1,k_2,k_3)
= -\frac65\fnl \[ P_\zeta(k_1)  P_\zeta(k_2) +\,{\rm cyclic} \] 
\label{fnldef} \,.
\ee
(Only the  term linear in $\fnl$ is
kept, which is justified  because the second term of \eq{corrchis}
is much less than the first term.) 
Current observation  \cite{komatsu,new} gives  $|\fnl|\lesssim 100$,
which makes the non-Gaussian fraction of $\zeta$ less
than $100\calp_\zeta\half\sim 10^{-3}$.
Absent a detection, PLANCK \cite{planck} will bring this down to roughly 
 $|\fnl|\lsim 1$.

Following 
Maldacena \cite{maldacena}, we define $\fnl$ by \eq{fnldef}
irrespectively of its origin, making it in 
 general  a  function of the wavenumbers.

The  trispectrum $T_g$ is   defined in terms of the 
connected four-point correlator by 
as
\be
\vev{g_{\bfk_1} g_{\bfk_2}g_{\bfk_3}g_{\bfk_4}}_c=
(2\pi)^3\delta^{(3)}(\bfk_1+\bfk_2+\bfk_3+\bfk_4) T_g
\,.
\ee
It is a function of six scalars,  defining  the 
quadrilateral formed by $\{\bfk_1,\bfk_2,\bfk_3,\bfk_4\}$.
If the curvature perturbation is given by \eqref{corrchis}, its
trispectrum to leading order is of the form  \cite{okamoto}
\be
T_\zeta =\frac12\tau\sub{NL} \[ P_\zeta(k_1) P_\zeta(k_2) 
P_\zeta(|k_{14}|) 
+\,23\,{\rm perms.} \]
\,,
\label{taudef}
\ee
with $\tau\sub{NL}=36\, f^2\sub{NL}/25$. In this expression, $k_{ij}\equiv
|\bfk_i+\bfk_j|$, and the permutations are of $\{\bfk_1,\bfk_2,\bfk_3,\bfk_4
\}$ giving actually 12 distinct terms. 

The authors of \cite{komatsu1,kunz} have measured
some connected terms of the angular trispectrum on the COBE DMR
data, with no significant detection of signals. 
Non-detection of the trispectrum by  the
COBE DMR data may be interpreted as $|\fnl|<10^4$ (95\% c.l.) based
upon theoretical expectations (Figure 3 of \cite{okamoto} with
$\ell\sub{max}=10$). Expressed as a bound on
$\tnl$, the current bound becomes $|\tnl| \lsim 10^{8}$.
When a bound on the trispectrum becomes available from 
 WMAP data, it might constrain non-gaussianity of the type \eqref{fnldef}
more strongly than the bispectrum. 
There is currently no estimate of the bound on the trispectrum which will
eventually be possible.

Following Maldacena's strategy for $\fnl$, we will define $\tau\sub{NL}$
by \eq{taudef} irrespectively of its origin, so that in general it depends
on the wave-vectors.  
In this Letter we will  calculate $\fnl$ and $\tau\sub{NL}$
in the case that the curvature perturbation has the form 
\eqref{uncorrchis}.

{\it The bispectrum.}~
We  take the spectrum $\calp_\sigma$ to be scale-invariant, which
will be a good approximation if  $\sigma$ is
sufficiently light. Since $\zetag$ and $\sigma$ are uncorrelated and the
curvature perturbation is dominated by $\zetag$, the spectrum
and the bispectrum of the curvature perturbation are
\bea
&&P_\zeta (k)=P_\zetag(k) +  P_\sigmas(k) \simeq  P_\zetag(k)
\label{phis1}\\
&&B_\zeta (k_1, k_2, k_3)=  B_{\sigma^2}(k_1, k_2, k_3)
\eea

The Fourier components of $\sigmas$ are given by
\be
(\sigmas)_\bfk = \frac1\tpq \int d^3q \sigma_\bfq \sigma_{\bfk-\bfq}
\label{phisk}
\,.
\ee
For non vanishing $\bfk$ and $\bfk'$
\bea
\vev{(\sigmas)_\bfk (\sigmas)_\bfkp} &=& \frac1\tps \int d^3p \,d^3p'\;
\vev{\sigma_\bfp \sigma_{\bfk-\bfp} \sigma_\bfpp \sigma_{\bfkp-\bfpp} }
\nonumber\\
&=& 2\delta^{(3)}(\bfk+\bfk')\int d^3p \, P_\sigma(p) P_\sigma(|\bfk-\bfp|)
\,\nonumber.
\eea
Taking $\calp_\sigma$ to be scale-independent \cite{myaxion}, 
\be
\calp_\sigmas(k) = \frac{k^3}{2\pi} \calp_\sigma^2 \int_\lmone 
\frac{d^3p}{ p^3 |\bfp-\bfk|^3 }
\label{psigs}
\,.
\ee
The subscript $\lmone$ indicates that the integrand 
is set equal to zero in a sphere of radius 
$\lmone$ around each singularity,  and  
the discussion makes sense only for $\lmone \ll k \ll \kmax$.
In this regime one finds \cite{myaxion} 
\be
\calp_\sigmas(k) = 4
\calp_\sigma^2 \ln(kL) \label{exact}
\,.
\ee
This expression gives
the correct variance
$\vev{\(\sigma^2-\vev{\sigmas}\)^2}=2\vev{\sigma^2}^2$, confirming
the consistency of the finite box approach.

Now consider the bispectrum $B_\zeta=B_\sigmas$.
Repeating the calculation leading to  \eq{psigs}, we find
eight terms  which can  brought into  a common
form by a redefinition of the integration variables, giving
\be
B_\zeta
={(2 \pi)^3}\calp_\sigma^3 \int_\lmone 
\frac{d^3p}{p^3 |\bfp-\bfk_1|^3 |\bfp+\bfk_2|^3}
\label{bis}\,,
\ee
where again $\lmone$ means that a sphere of radius $\lmone$ is cut out around
each singularity.
The integral may be estimated by adding the contributions of the three
singularities, and comparing the result with \eq{fnldef} one finds
\be
\fnl = -A \frac{20}{3} \frac{\calp_\sigma^3}{\calp_{\zeta}^2} \ln(kL)
\,.
\ee
with $A=O(1)$ and $k=\min k_i$. Expressed in terms of the 
non-gaussian fraction 
$r_\zetas\equiv  (\calp_\sigmas/\calp_\zeta)\half$, this becomes
$r_\zetas\simeq 0.04 \fnl\third$

To compare with observations,  we   choose a  minimal box size, so
that $\ln(kL)$ is roughly of order 1 on cosmological scales.
Since $\fnl$ has only mild scale-dependence, we can use  observational
bounds on that are obtained by taking $\fnl$ to be scale-independent.
The current bound $|\fnl|\lsim 100$ corresponds to 
$r_\sigmas\lesssim 0.18$. Planck \cite{planck} will be able to
detect an un-correlated Gaussian-squared component of the
primordial curvature perturbation, provided that its relative
amplitude $r_\sigmas$ is bigger than $0.04$.

{\it The trispectrum.}~ 
Computing the connected four-point correlator of $\sigma^2$ we find
\bea
T_\zeta &=& 4\pi^5 \calp_\sigma^4 \int_\lmone \frac{ d^3p}
{p^3 |\bfp-\bfk_1|^3 |\bfp+\bfk_2|^3|\bfp+\bfk_{24}|} \nonumber \\
&+&\,23\,{\rm perms} \label{Tzeta}
\,,
\eea
where the permutations are of $\{\bfk_1\bfk_2\bfk_3\bfk_4\}$
giving actually 3 distinct terms.
A sphere of radius $\lmone$ is cut out around each singularity,
and imposing $k\gg\lmone$ on the  wavenumbers in \eq{phisk}
gives  $|\bfk_{ij}|\gg\lmone$. 
The  correspondence between \eqs{Tzeta}{taudef} is seen by
drawing Feynman-like diagrams as for instance in \cite{renorm}.
Estimating the integral by adding the contributions  of the four 
singularities, and comparing with \eq{taudef},  one finds
\be
\tau\sub{NL}= 16 B {\calp^4_\sigma\over\calp^3_\zeta } \,\ln(kL)
\,,
\ee
with $B=O(1)$ and $k=\min\{k_i,|\bfk_{jm}|\}$. For the fraction of 
non-gaussianity this corresponds to $r_\sigmas\sim 10^{-1} \tnl\quarter$.
This result might be competitive with  the one coming from $\fnl$, when 
 $\tnl$ is bounded by  WMAP. 

{\it A Gaussian-squared isocurvature perturbation.}~In addition to
the primordial curvature perturbation, there might be a primordial
isocurvature perturbation. Like the former, the latter is to be
evaluated a few Hubble times before cosmological scales start to
enter the horizon, and is specified by
\be S_i = \frac{\delta n_i}{n_i} -  \frac{\delta
n_\gamma}{n_\gamma} \,,
\ee
where the $n$'s are number densities
and $i$ runs over  cold dark matter particles ($C$), baryonic
matter ($B$) and neutrinos ($\nu$). Observation requires roughly
$|S_i|\lesssim 0.1\zeta$. Depending on its origin, $S_i(\bfx)$
might be a constant multiple of $\zeta(\bfx)$, but it might
instead be uncorrelated with it \cite{lw2}. The classic example of the latter
case is  a CDM isocurvature perturbation  caused by a
perturbation in the axion field. In that situation $S_C$ will be
the sum of terms linear and  quadratic in the axion field
\cite{myaxion}. Let us suppose that the latter dominates and that $\zeta$
is Gaussian. On
large angular scales, ignoring baryons, the temperature
anisotropies are given in term of the primordial curvature and
isocurvature perturbation through
\be \(\frac{\Delta T}{T}\)\sub{SW}=\(\frac{\Delta T}{T}\)\sub{A} +
\(\frac{\Delta T}{T}\)\sub{S}= -\frac{\zeta}{5} - \frac25
S_C \ee The angle-averaged bispectrum defined e.g. in
\cite{spergel} will consist only of the isocurvature one i.e.
$B_{\ell_1, \ell_2, \ell_3}=B^{(S_C)}_{\ell_1, \ell_2, \ell_3}$.
Since the coefficients of $\zeta$ and $S_C$ are similar, the 
observational bound on $\calp_{S_C}/\calp_\zeta$ from non-Gaussianity
will be similar to the
one on $\calp_\sigmas/\calp_\zeta$, and hence competitive with the existing
bound. 

{\it The logarithmic scale dependence.}~We have found that the 
spectrum, bispectrum and trispectrum all increase like $\ln(kL)$
with $k$ a representative wavenumber. 
This factor arises from the infrared divergence of \eq{phisk},
which in turn is due to the interference of very long wavelength components.
We  adopted a minimal box size making
the factor of order one  on cosmological scales, though it 
might still be significant on  very small
scales relevant for instance for primordial black hole formation.
But irrespective of practicalities, the appearance of the factor 
 seems to contradict a basic tenet of  physics concerning the 
use of
Fourier series; that the box size should be irrelevant if it is much bigger
than the scale of interest. 

To see what is going on, we need to consider
the generic perturbation \eqref{withlin} which contains a linear term
$a\sigma$. The crucial point is that the mean $\vev{\sigma}$ is supposed to
vanish {\em within the chosen box} of size $L$. Suppose now that we go to
box of size $M\ll L$, still comfortably surrounding the region of 
observational
interest. (It might be the whole observable Universe, or a much smaller
 region around us where say the distribution of primordial black holes has
been detected.) 
Inside the small box the appropriate variable is
 $\hat\sigma\equiv \sigma - \sigmam$, and we will denote expectation values
inside the small box by $\vev{}_M$. Then
\be
\zetasig(\bfx) = a(M) \hat\sigma(\bfx) + \( \hat\sigma^2(\bfx)-
\vev{\hat\sigma^2}\)
\label{uncorr2}
\,,
\ee
with $a(M) = a + 2\vev{\sigma}_M$.
Finally, instead of locating the small box around a particular observed 
region, let its location vary so that
 $\vev{\sigma}_M$ becomes the original perturbation $\sigma$
smoothed on the scale $M$. It is  clear that the stochastic properties
of the perturbations within a randomly-located small box will be the same,
whether we use such a box directly or whether we calculate them using
instead the big box. 

Let us see how  this works for the spectrum, bispectrum and trispectrum
of $\zetasig$. The spectrum is 
$\calp_\zetas = a^2 \calp_\sigma + \calp_\sigmas $.
Defined within a particular small box it becomes
\be
\calp_\zetas(M) = a^2(M) \calp_\sigma + \calp_\sigmas
\,.
\ee
Using \eq{exact},  the expectation value is
\be
\vev{\calp_\zeta(M)} = 4 \calp_\sigma^2 \( \ln(L/M) + \ln(kL) \) 
= \calp_\zeta
\,.
\ee
These equations translate to the following words;
going to the small box increases the  the size of the
cutoff  around the singularities, from $\lmone$ to $M^{-1}$,  but this
is compensated by the change in the coefficient $a(M)$.

Exactly the same thing happens for the bispectrum and trispectrum.
Using our previous results, the bispectrum within the large box is 
\bea
B_\zeta &=& 8\pi^4 a^2 \calp_\sigma^2 
\( \frac1{k_1^3k_2^3} +\,{\rm cyclic} \) \nonumber \\  
&+& (2\pi)^3 \calp_\sigma^3 \int_\lmone 
\frac{d^3p}{p^3 |\bfk_1-\bfp|^3 |\bfk_2+\bfp|^3 }
\,.
\eea
Replacing $a$ by $a(M)$ and $L$ by $M$ gives the 
 bispectrum $B_\zeta(M)$ within  a small box, leading to
\bea
\vev{B_\zeta(M)} &=& (2\pi)^3 \calp_\zeta^3 
\[ 4\pi \ln(L/M)  \( \frac1{k_1^3k_2^3}  +\,{\rm cyclic} \) 
\right. \nonumber \\
&+& \left. \int_{M^{-1}} \frac{d^3p}{p^3 |\bfk_1-\bfp|^3 |\bfk_2+\bfp|^3 }
 \]
\,.
\eea
The  first term changes $\int_{M^{-1}}$ to $\int_\lmone$,  and a 
similar calculation works for the trispectrum.

What we are finding here is a logarithmic `running' of the (typical value of)
the parameter $a$ 
with the box size,  
similar to the running of field theory parameters  with the 
renormalization scale. The logarithm would become a power if 
 $\calp_\sigma(k)$ went like a positive power. If more fields and/or
higher terms in the expansion of $\zeta$ 
were allowed,  we would arrive at 
`renormalization group' equations relating the running of the coefficients
in the expansion. That is hardly of practical interest for the primordial
curvature perturbation, but it will be applicable when the formalism developed
in \cite{renorm} is extended to include non-Gaussianity.

Returning to the primordial perturbation, consider again  our assumption that 
$\zeta$ (or $S_i$) is  quadratic in $\sigma$, corresponding to $a\simeq 0$. 
If this  assumption is valid within a box 
exponentially larger than the observable Universe, then it can be valid
also for the minimal box that we adopted only if our location is to 
some extent untypical. This  situation was noticed for the 
axion \cite{lindeaxion,myaxion}, providing an early example of a 
theoretically-motivated anthropic consideration.

 {\it Conclusion.}~We have considered the
effect of a sub-dominant primordial perturbation, which is uncorrelated
with the main component. Unlike the main component, this one may be highly
non-Gaussian, with the result that its stochastic properties are quite 
different from those of the main component. They have to be defined with
respect to a comoving box of finite size, but a change in the box size is
compensated by a change in the coefficient of the Gaussian component 
as measured by a typical observer, so that
physical results are unchanged for such an observer. 

Coming to the observational side, 
and assuming
that the sub-dominant is the square of a Gaussian,  it will eventually be
detectable if 
it contributes more than a few percent of the total. A smoking gun
for this setup would be the detection of both primordial
gravitational waves (indicating that the inflaton gives the
dominant contribution to the primordial curvature perturbation)
and non-Gaussianity (indicating that this contribution cannot be
the only one). Another smoking gun would be provided by the
observation of both the bispectrum and the trispectrum, the latter being at 
least equally important.

{\it Acknowledgments.}~ We  thank Rob Crittenden and
Misao Sasaki for valuable input in the early stages of this work,
 Nicola Bartolo for reminding us of reference
\cite{Fan:1992wv}, and  Eiichiro Komatsu for very useful
correspondence regarding the trispectrum. This work is supported
by PPARC grants PPA/G/O/2002/00469, PPA/V/S/2003/00104,
PPA/G/O/2002/00098 and PPA/S/2002/00272 and EU grant 
MRTN-CT-2004-503369.


\begin{thebibliography}{99}


\bibitem{treview} 
A.~R.~Liddle and D.~H.~Lyth, {\it Cosmological inflation and
large-scale structure}, Cambridge University Press, 2000.

\bibitem{sl}
D.~Seery and J.~E.~Lidsey,
arXiv:astro-ph/0506056.

\bibitem{maldacena}
J.~Maldacena,
JHEP {\bf 0305}, 013 (2003)

\bibitem{curvaton}
D.~H.~Lyth and D.~Wands,
Phys.\ Lett.\ B {\bf 524}, 5 (2002);
T.~Moroi and T.~Takahashi,
Phys.\ Lett.\ B {\bf 522}, 215 (2001) (Erratum-ibid. B {\bf 539}, 303 (2002)).

\bibitem{also}
S.~Mollerach,
Phys.\ Rev.\ D {\bf 42}, 313 (1990);
A.D.~Linde and V.~Mukhanov,
Phys.\ Rev.\ D {\bf 56}, 535 (1997);
 K.~Enqvist and M.~S.~Sloth,
  Nucl.\ Phys.\ B {\bf 626}, 395 (2002)

\bibitem{reheating}
G.~Dvali, A.~Gruzinov and M.~Zaldarriaga,
Phys.\ Rev.\ D {\bf 69}, 023505 (2004) 
L.~Kofman,
arXiv:astro-ph/0303614.

\bibitem{lr2}  D.~H.~Lyth and Y.~Rodriguez, astro-ph/0504045.


\bibitem{sasaki1}
D.~H.~Lyth, K.~A.~Malik and M.~Sasaki,
JCAP {\bf 0505}, 004 (2005)


\bibitem{spergel}
E.~Komatsu and D.~N.~Spergel, Phys. Rev. D {\bf 63}, 063002 (2001).


\bibitem{komatsu} E.~Komatsu {\it et al.}, Astrophys. J. Suppl. Ser.
{\bf 148}\, 119 (2003).




\bibitem{Langlois:2004nn}
D.~Langlois and F.~Vernizzi,
Phys.\ Rev.\ D {\bf 70}, 063522 (2004) [arXiv:astro-ph/0403258].

\bibitem{ev}
  K.~Enqvist and A.~Vaihkonen,
  JCAP {\bf 0409}, 006 (2004)



\bibitem{Fan:1992wv}
Z.~H.~Fan and J.~M.~Bardeen,
preprint UW-PT-92-11. unpublished.



\bibitem{trispectrum}
W.~Hu, Phys. Rev. D {\bf 64}, 083005 (2001).

\bibitem{okamoto}
T.~Okamoto and W.~Hu,
Phys.\ Rev.\ D {\bf 66}, 063008 (2002)

\bibitem{myaxion}
D.~H.~Lyth,
Phys.\ Rev.\ D {\bf 45}, 3394 (1992).

\bibitem{planck} {\tt http://planck.esa.int/}

\bibitem{renorm}
  M.~Crocce and R.~Scoccimarro,
  arXiv:astro-ph/0509418.




\bibitem{komatsu1}
E.~Komatsu,
arXiv:astro-ph/0206039.


\bibitem{kunz} M. Kunz, A. J. Banday, P. G. Castro, P. G. Ferreira and
K. M. G\'orski, Astrophys. J. Lett. 563, L99 (2001).


\bibitem{new}  P.~Creminelli, A.~Nicolis, L.~Senatore, M.~Tegmark and M.~Zaldarriaga,
  arXiv:astro-ph/0509029.


\bibitem{lw2}
 D.~H.~Lyth and D.~Wands,
  Phys.\ Rev.\ D {\bf 68}, 103516 (2003)


\bibitem{double}
 A.~A.~Starobinsky,
  JETP Lett.\  {\bf 42}, 152 (1985)
  [Pisma Zh.\ Eksp.\ Teor.\ Fiz.\  {\bf 42}, 124 (1985)].

\bibitem{preheating}
 M.~Bastero-Gil, V.~Di Clemente and S.~F.~King,
  Phys.\ Rev.\ D {\bf 70}, 023501 (2004);
  E.~W.~Kolb, A.~Riotto and A.~Vallinotto,
  Phys.\ Rev.\ D {\bf 71}, 043513 (2005).

\bibitem{lindeaxion}
 A.~D.~Linde,
  Phys.\ Lett.\ B {\bf 201}, 437 (1988).

\end{thebibliography}
\end{document}